\documentclass[12pt,preprint]{aastex}

\slugcomment{To be submitted to Astrophys. J.}

\shorttitle{Emergence of Hyperons in Black Hole Formation}
\shortauthors{Sumiyoshi et al.}

\begin{document}


\title{Emergence of hyperons in failed supernovae:
trigger of the black hole formation}


\author{K. Sumiyoshi}
\affil{Numazu College of Technology, 
       Ooka 3600, Numazu, Shizuoka 410-8501, Japan}
\email{sumi@numazu-ct.ac.jp}

\author{C. Ishizuka}
\affil{Department of Cosmosciences, Graduate School of Science, 
       Hokkaido University, Sapporo 060-0810, Japan}
\email{chikako@nucl.sci.hokudai.ac.jp}

\author{A. Ohnishi}
\affil{Yukawa Institute for Theoretical Physics, Kyoto University, 
       Kyoto 606-8502, Japan}
\email{ohnishi@yukawa.kyoto-u.ac.jp}

\author{S. Yamada}
\affil{Science and Engineering
       \& 
       Advanced Research Institute for Science and Engineering, \\
       Waseda University, 
       Okubo, 3-4-1, Shinjuku, Tokyo 169-8555, Japan}
\email{shoichi@heap.phys.waseda.ac.jp}

\and

\author{H. Suzuki}
\affil{Faculty of Science and Technology, Tokyo University of Science,
       Yamazaki 2641, Noda, Chiba 278-8510, Japan}
\email{suzukih@ph.noda.tus.ac.jp}



\begin{abstract}
We investigate the emergence of strange baryons 
in the dynamical collapse of a non-rotating massive star to a black hole 
by the neutrino-radiation hydrodynamical simulations in general 
relativity.  
By following the dynamical formation and collapse of nascent proto-neutron star 
from the gravitational collapse of a 40M$_{\odot}$ star 
adopting a new hyperonic EOS table, 
we show that the hyperons do not appear at the core bounce  
but populate quickly at $\sim$0.5--0.7 s after the bounce 
to trigger the re-collapse to a black hole.  
They start to show up off center owing to high temperatures 
and later prevail at center when the central density becomes 
high enough.  
The neutrino emission from the accreting proto-neutron star 
with the hyperonic EOS stops much earlier than the corresponding 
case with a nucleonic EOS while the average energies 
and luminosities are quite similar between them.  
These features of neutrino signal are a potential probe 
of the emergence of new degrees of freedom 
inside the black hole forming collapse.  
\end{abstract}





\section{Introduction}\label{sec:intro}
The fate of gravitational collapse of massive stars 
in the mass range of $\gtrsim$20M$_{\odot}$ attracts much 
attention recently in the attempt to reveal the origin of
diversity of supernovae \citep{nom07}.  
Their collapse will lead to the formation of black hole 
and provide us with the test bed for the appearance of 
exotic particles at high density/temperature.  

There appear to be at least two types of massive stellar death in this 
mass range.  Faint supernovae (or failed supernovae) are one of
them, having very low explosion energies 
as observed in SN 1997D and SN 1999br \citep{nom07}.  
In contrast to the other branch referred to sometimes as the
hypernova branch because of their high explosion energies, 
their progenitors are supposed to be slowly rotating 
and collapse to a black hole with no bright optical display.  
Such a quiet death may consist of a significant fraction 
($\sim 30 \%$) of all massive stellar collapses and 
may be detected in the near future, for example, by 
the planned survey of disappearance of 
supergiants \citep{koc08}.  

Recent studies by \citet{sum06,sum07} demonstrated that 
the neutrino burst from these quiet collapses of massive stars 
can be used not only as a hallmark of black hole formation but
also as a probe of dense matter.  In fact, 
the profile of neutrino burst emitted 
from the collapsing proto-neutron star born temporarily 
in the failed explosion 
has unique characteristics to distinguish them 
from ordinary supernova neutrinos \citep[See also][]{bur88,lie04}.  
The duration of neutrino burst is short ($\sim$1 s) 
because the neutrino emission is terminated 
shortly after the black hole formation via the re-collapse 
of proto-neutron star with intense accretions.  
The energies and luminosities of neutrinos increase quickly 
thanks to the rapid contraction of accreting proto-neutron star.  
These characteristics have been used to search for the black hole 
formation in the archive of terrestrial neutrino detector \citep{ike07}
with no positive detection so far.  

The future observation of short neutrino bursts, however, 
would constrain the stiffness of equation of state (EOS) 
and the appearance of new particles, 
on which we focus in the current study.  
The EOS crucially determines when the critical mass is reached by 
an ever fattening proto-neutron star 
and, therefore, the duration of neutrino burst.  
It should be noted that the previous works \citep{sum06,sum07} took
into account only the nucleonic degree of freedom. If  
new degrees of freedom such as hyperons (series of strange baryons) 
appear at some point of the evolution owing to high densities or 
temperatures, however, then the EOS will become softer \citep{gle00},  
triggering the re-collapse of proto-neutron star earlier, and 
the neutrino burst may become even shorter.  

In this Letter, we report the first serious numerical study of 
the dynamical collapse of a non-rotating massive star 
adopting a new EOS with hyperons \citep{ish08}.  
We reveal when the hyperons appear during the stages from 
the initial collapse of progenitor stars, core bounce and 
proto-neutron star evolution until the black hole formation.  
By the neutrino-radiation hydrodynamical simulations, 
we demonstrate the outcome of hyperon emergence 
in the dynamics and the neutrino emission.  
The hyperonic EOS we adopt here is based on the same framework 
as the nucleonic EOS by \citet{she98a,she98b}.  
As a natural extension of the former, 
the latter enables us to extract clearly the effect of 
hyperons.  


We stress that 
the current scenario of black hole formation 
is entirely different from the delayed collapse 
of meta-stable proto-neutron stars 
and the associated termination of neutrino signals 
in $\sim$1--100 s \citep{kei95,pon99,bau96b}.  
In fact, the proto-neutron star mass is fixed in successful explosions 
while it increases continuously and rapidly by the accretion in the 
failed explosion considered in the current study.  
The rapid appearance of hyperons in the {\it dynamical} collapse 
accompanied by a short neutrino burst is, therefore, clearly different 
from the characteristics seen in the {\it quasi-static} 
cooling of proto-neutron stars \citep{pon99,pon01b,pon01a}, 
in which the emergence of new compositions (hyperons or any other forms) 
occurs much later when the matter becomes neutron-rich through 
the deleptonization except for rather extreme choices of interactions.  
Accordingly, the effect of hyperon mixture on the neutrino signal 
manifests itself only in the exponential tail of time evolution 
and hence is hard to detect in observations.  

%


%
%


\section{Hyperon equation of state}\label{sec:hyper-eos}

We employ the sets of EOS both with and without the hyperons 
to explore their influence.  
As a reference of nucleonic EOS, 
we adopt the Shen EOS \citep{she98a,she98b}, 
which describes the mixture of neutrons, protons, 
alpha particles and nuclei in the relativistic mean 
field theory.  
The hyperonic EOS we adopt here was recently developed 
by \citet{ish08}, extending the Shen EOS to SU(3) symmetry 
to include the octet baryons.\footnote{The tables 
of the hyperonic EOS are available for public use.}
It should be stressed that the nucleon sector is not changed 
from the Shen EOS and, as a result, the low density 
($\le10^{14}$ g/cm$^3$) part of the hyperonic EOS is connected
smoothly with the Shen EOS. The interactions of hyperons are
determined by the recent experimental data of hypernuclei.  
As for the $\Sigma^{-}$-potential in symmetric nuclear matter in particular, 
we choose the repulsive value (+30 MeV), which is preferred lately.  
The maximum masses of cold neutron stars 
by the Shen EOS and the hyperonic EOS are 
2.2M$_{\odot}$ and 1.6M$_{\odot}$, respectively.  
The latter value does not change much even for 
other choices of interactions and/or  the inclusion of thermal pions \citep{ish08}.  
The other sets of the hyperonic EOS, therefore, are expected to 
give the results not much different from the current ones.  

In order to discuss the dependence on the EOS 
within the nucleonic degree of freedom, 
we also utilize the profiles of neutrino bursts 
previously obtained by the Lattimer-Swesty EOS \citep{lat91} 
with the incompressibility of 180 MeV, which 
is softer than the Shen EOS.  
Its maximum neutron star mass is 1.8M$_{\odot}$.  
%
Although other new degrees of freedom such as quarks 
may appear near the black hole formation, we concentrate here 
on the hyperon mixture and refer readers to \citet{nak08}, which 
suggests that the quark mixture may occur only 
at the last moment of black hole formation. Needless to say, 
it is important to describe the phase transitions 
from hyperonic matter to quark matter as well as 
the meson-condensations consistently \citep{gle00}.  
Such a task is being undertaken currently 
to implement in the numerical simulations.  

\section{Numerical Simulations}\label{sec:simulations}

We perform the numerical simulations of neutrino-radiation 
hydrodynamics in general relativity under the spherical symmetry 
in the same manner as \citet{sum06,sum07}.  
We follow the dynamics from the onset of gravitational collapse 
of a progenitor star, 
through the core bounce and the post-bounce evolution of compact objects 
by the accretion of outer layers, up to 
the formation of the apparent horizon \citep{nak06}.  
As an initial model, we employ the central part of the progenitor 
model of a 40M$_{\odot}$ star by \citet{woo95}.  
We refer to the model using the Shen EOS as model SH and the 
corresponding model with the Lattimer-Swesty EOS as model LS.  
We name the new model 
using the Ishizuka EOS as model IS, which we report mainly in this Letter.  
For the detailed information on the numerical settings and results 
of models SH and LS, we refer to \citet{sum07}.  

We adopt the conventional weak interaction rates 
as in the previous numerical simulations \citep{sum05}.  
In the current study, we ignore the neutrino-hyperon interactions
entirely. This is not so bad an approximation as it seems, however. 
In fact, owing to the hyper-accretion of outer layers, the neutrino
emissions occur dominantly near the surface of
the proto-neutron star, where hyperons are scarce. 
Moreover, since hyperons appear only 
for $\sim$200 ms before the black hole formation, 
the possible modifications of diffusion rates in the central region,
where hyperons are mostly populated, do not have enough time to affect 
the neutrino fluxes. This reflects the sharp contrast in the durations
of neutrino emissions $\sim$1 s in the present case and $\sim$20 s for 
the quasi-static proto-neutron star cooling.  

\section{Numerical results}\label{sec:results}

\begin{figure}
\epsscale{1}
\plotone{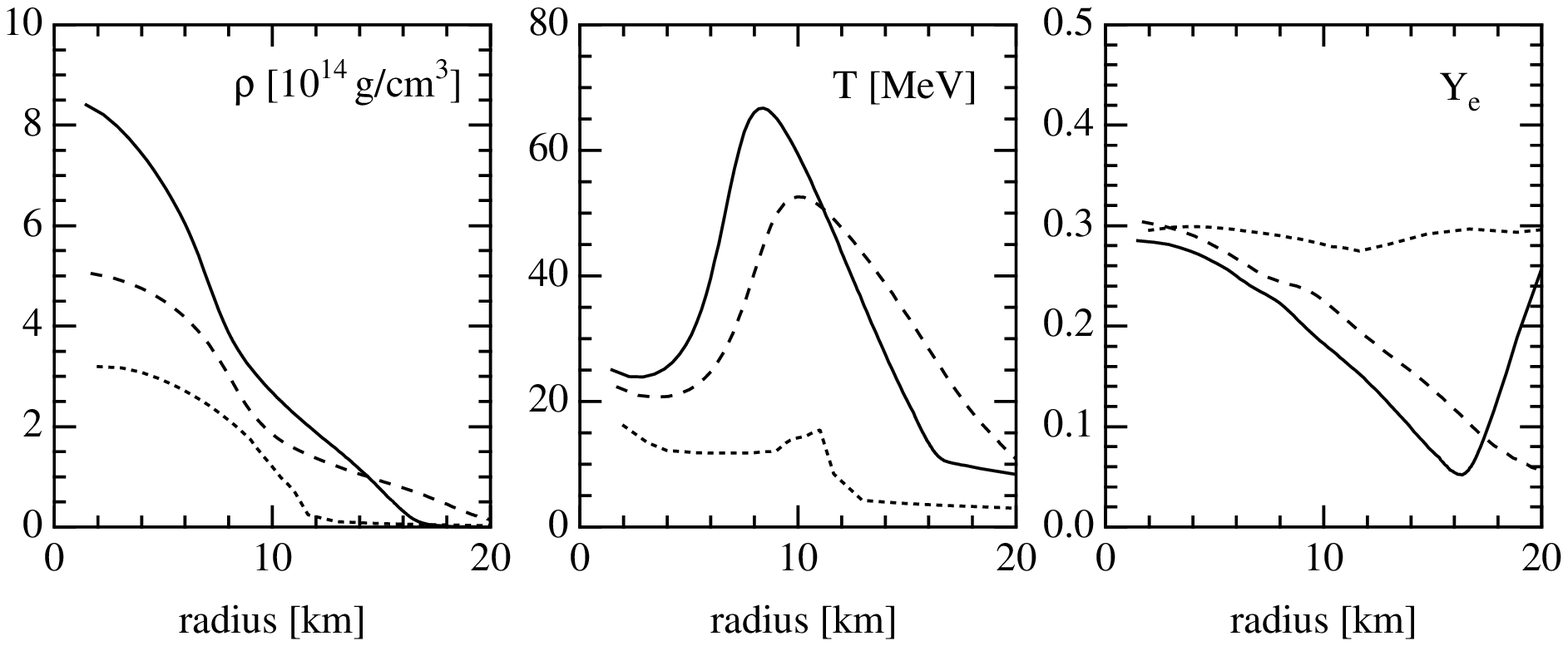}
\caption{Density (left), temperature (center) and 
electron fraction (right) profiles for IS 
at t$_{pb}$=0, 500 and 680 ms are shown 
by dotted, dashed and solid curves, respectively, 
as a function of radius.}
\label{fig:profile}
\end{figure}

It is of particular interest to see when hyperons first appear 
in the collapse of massive stars, since this is the first quantitative
investigation utilizing a realistic dynamics and hyperonic EOS.  
We found that hyperons do not appear at the core bounce, 
since the central density and temperature are simply not high enough 
as shown in Fig.~\ref{fig:profile}.  In fact, 
the central density at the core bounce is just above 
the nuclear matter density. Accordingly, the core bounce and the
launch of shock wave followed by the recession, occur exactly in the
same way as in model SH. The proto-neutron star without hyperons 
is born by t$_{pb}$=300 ms, where t$_{pb}$ denotes the time 
after the bounce.

As the density and temperature increase in the contracting 
proto-neutron star with the growing mass by the accretion, hyperons 
first appear off center at t$_{pb}$$\sim$500 ms.  
They soon populate in substantial amount at center 
and become major compositions by t$_{pb}$=680 ms, 
at which the central density 
exceeds three times the nuclear matter density 
as seen in Fig.~\ref{fig:profile}.  
This quick appearance of hyperons 
in the black hole forming collapse is different 
from what happens in non-accreting 
proto-neutron stars, where hyperons appear 
gradually over $\gtrsim$10 s \citep{pon99} through the cooling and deleptonization.  
The dynamical collapse to black hole ensues immediately 
when the proto-neutron star mass reaches its critical value 
owing to the softening of EOS by the production of hyperons.  
The black hole is formed at t$_{pb}$=682 ms in model IS, 
which is much earlier than t$_{pb}$=1345 ms in model SH.

\begin{figure}
\epsscale{1.16}
\plottwo{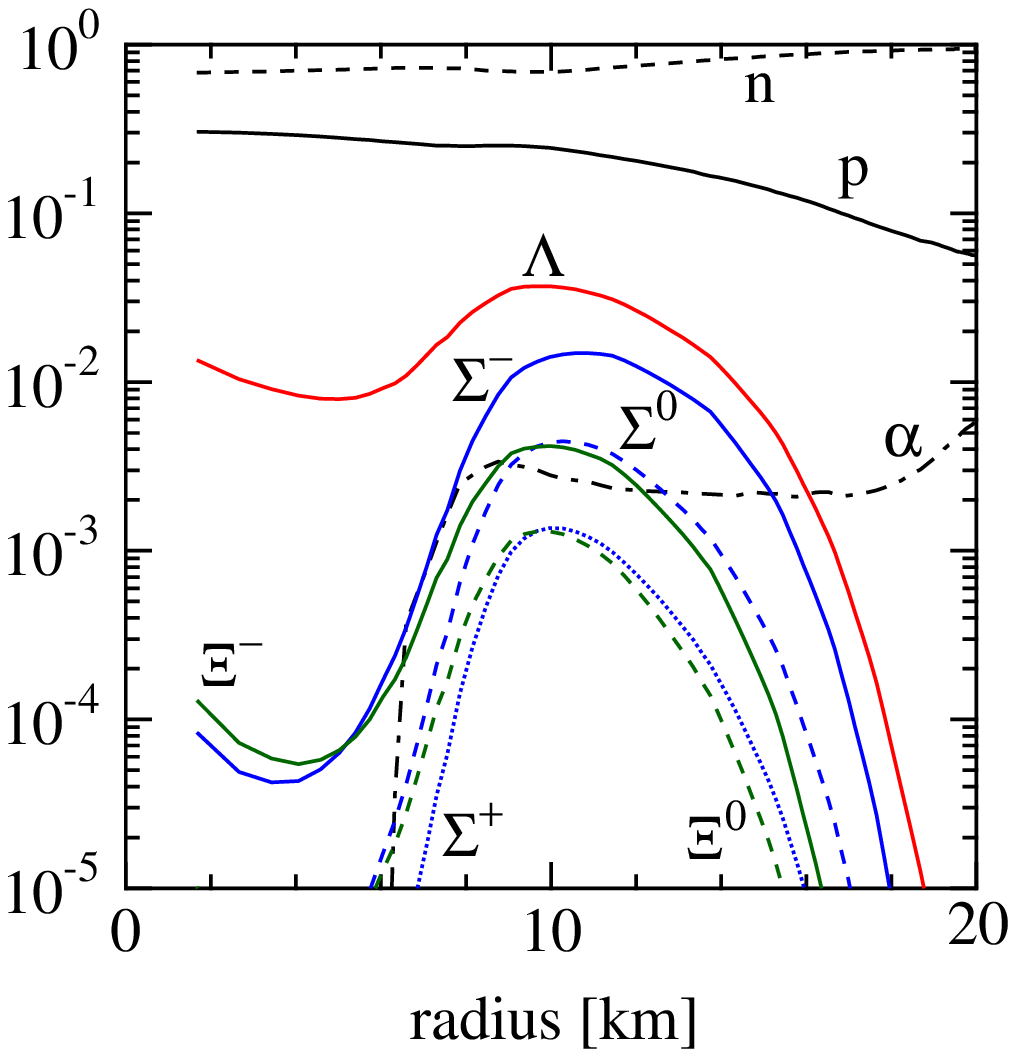}{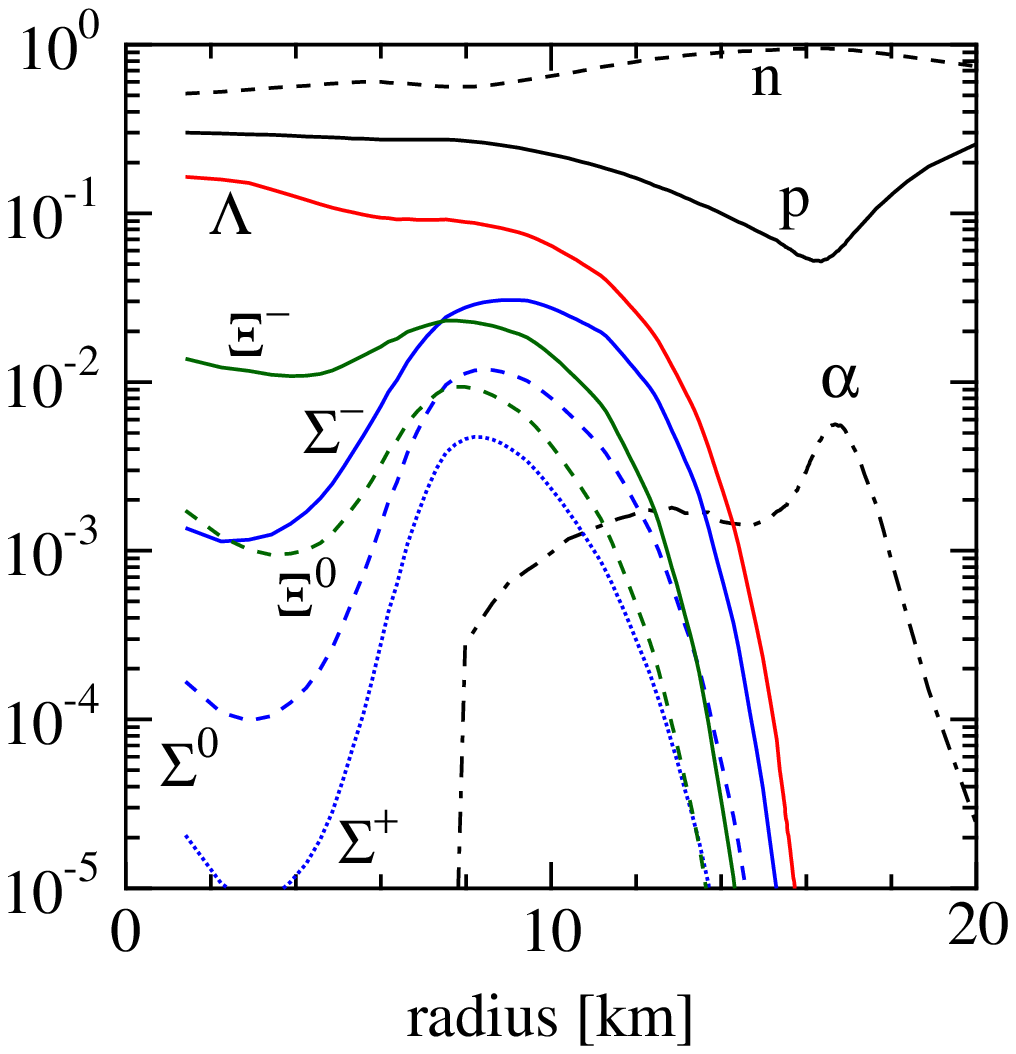}
\caption{Mass fractions of hyperons in model IS 
are shown as a function of radius 
at t$_{pb}$=500 (left) and 680 ms (right).}
\label{fig:composition}
\end{figure}

We show two snapshots of the compositions inside the proto-neutron star 
in Fig.~\ref{fig:composition}.  
At t$_{pb}$=500 ms (left panel), the mass fraction of hyperons, 
mostly $\Lambda$ and $\Sigma^{-}$ particles, is a few $\%$ 
at the maximum around 10 km (0.7M$_{\odot}$ in mass coordinate).  
The abundance of hyperons becomes significant at t$_{pb}$=680 ms 
(right panel), 
since the central density becomes high enough by this time.  
$\Lambda$ particles exist as a major composition ($\ge 10 \%$) 
and $\Xi^{-}$ particles are the next abundant one.  
$\Sigma^{-}$ particles are suppressed at the central region 
because of the repulsive potential we have chosen.  
This hierarchy is different from that for 
attractive potentials, where negatively-charged $\Sigma^{-}$ particles
are favored \citep{pon99}.  

It is remarkable that hyperons appear 
initially off center owing to high temperatures.  
Although the density at 10 km is not so high at t$_{pb}$=500 ms, 
the temperature exceeds 50 MeV at the peak 
(See Fig.~\ref{fig:profile}), 
where the shock wave is produced and the entropy is generated.  
The conversion to hyperons occurs 
through the high energy tail above the hyperon mass of thermal nucleons.  
This temperature effect keeps the fraction 
of hyperons around 10 km large even at late stages.  
The appearance of 
other particles or phase transitions may also occur owing to the same temperature effect
and lead to similar consequences.  
It is also interesting to note that the hyperon emergence 
is not controlled by the deleptonization as in the 
proto-neutron star cooling but driven by the rapid increase 
of density and temperature in the contracting proto-neutron star.  
The electron fraction at center remains high ($\sim$0.3) 
up to the end whereas it decreases at outer part 
due to neutronization as seen in Fig.~\ref{fig:profile}.  
%
%

\begin{figure}
\epsscale{1.0}
\plotone{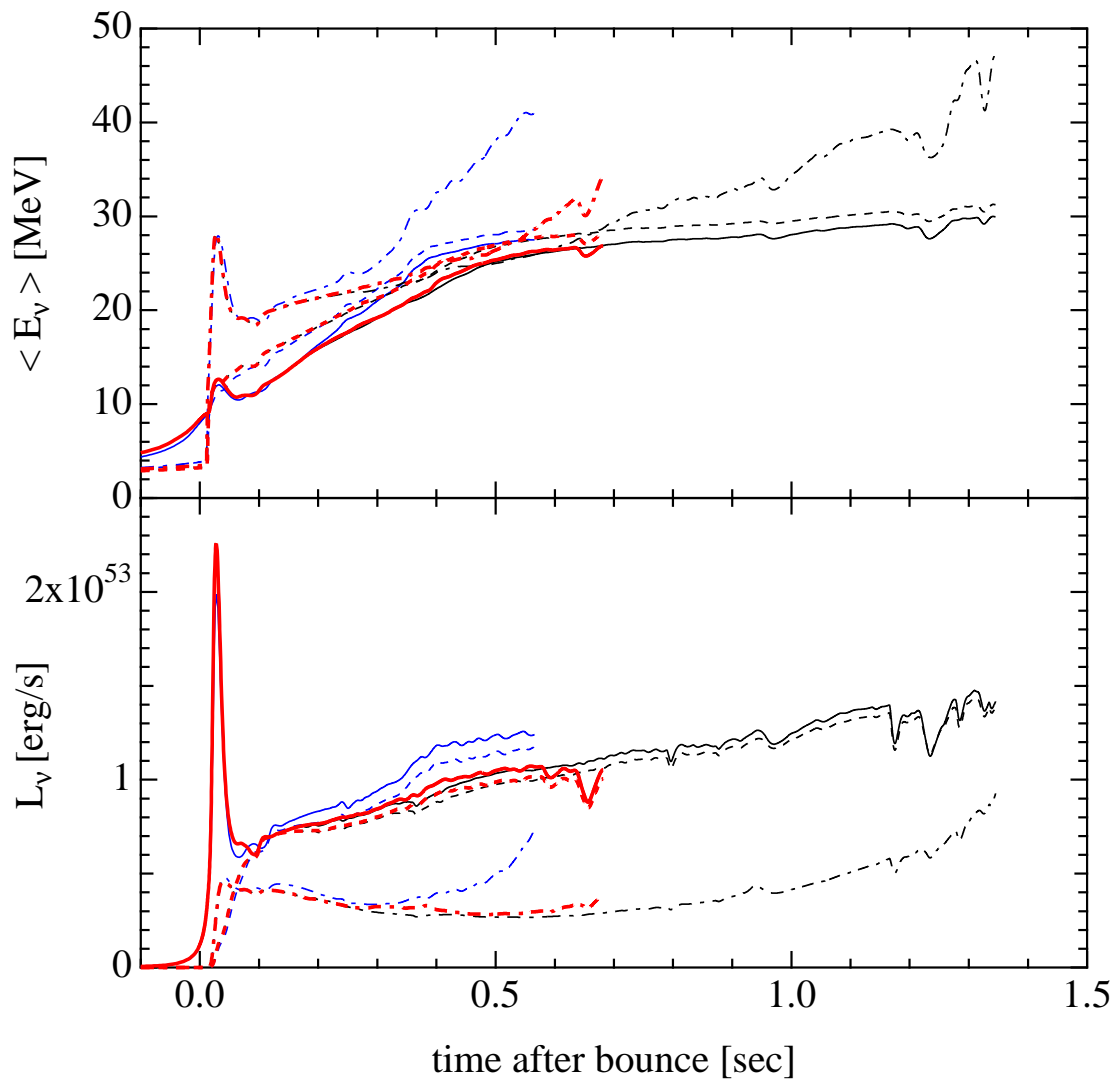}
\caption{Average energies and luminosities 
of $\nu_e$ (solid), $\bar{\nu}_e$ (dashed) and $\nu_{\mu/\tau}$ (dash-dotted) 
for model IS are shown as a function of time after bounce.  
The results for model SH and LS are shown by thin lines with the same notation.  }
\label{fig:neutrino}
\end{figure}

Whether the emergence of hyperons described above can be 
probed by the neutrino signal is a matter of great concern.  
We found that the duration of neutrino burst in model IS 
is clearly shorter than that in model SH 
as a result of the earlier black hole formation.  
We display the time profiles of average energies and 
luminosities of neutrinos for models IS and SH 
in Fig.~\ref{fig:neutrino}.  
It is apparent that these quantities are quite similar between 
two models\footnote{The slight oscillations in the curves 
occur owing to insufficient numerical resolutions 
at the final phase of computations.  } 
and only the duration is different.  
This similarity in neutrino emissions is a direct consequence of the 
almost identical accretion history irrespective of the presence of 
hyperons inside the proto-neutron star. In fact, we found that 
the mass of the proto-neutron star in model IS 
increases exactly in the same way as in model SH 
as a function of time.  
The critical (gravitational) mass, 2.1M$_{\odot}$, is reached 
earlier in model IS, which is smaller than 2.4M$_{\odot}$ in model SH.  
We remark here that these critical masses are for the hot proto-neutron 
stars with neutrinos, being different from those for the cold and 
neutrino-less neutron stars.  
It is noticeable that 
the average energy of $\nu_{\mu/\tau}$ increases 
faster in model IS than in model SH after t$_{pb}$=500 ms.  
Because of the hyperon emergence, the contraction of proto-neutron 
star is accelerated and leads to the quicker temperature increase. 

The intense and short neutrino burst may be used to infer the 
properties of the hyperonic matter.  
Since the hyperon appearance in the proto-neutron star triggers its 
earlier re-collapse to black hole, we would be able to estimate
from the duration of neutrino emissions the critical mass for the 
conversion to the hyperonic matter. Exrta information on the
progenitor would be highly helpful \citep{sum08} and will be plausible.  
It should be noted, however, that new degrees of freedom 
other than hyperons may also result in similar neutrino bursts.  
The further complication may arise from the uncertainties in the 
nucleonic EOS. We show the neutrino emissions for model LS in 
Fig.~\ref{fig:neutrino} to elucidate this. The softer LS EOS terminates
the neutrino burst at t$_{pb}$=566 ms without hyperons and gives 
a more rapid rise of the luminosities and energies  
for all flavors of neutrinos.  
It is hence necessary to look into more detailed differences 
of spectra in order to disentangle the degeneracy.  
The differences in $\nu_{\mu/\tau}$ will be important 
to distinguish them from one another 
when neutrino oscillations are taken into account.  

\section{Summary and Discussions}\label{sec:summary}

We demonstrated that the hyperon emergence 
in the collapse of a non-rotating massive star 
will produce an intense but short neutrino burst 
that may be used as a probe into the hyperonic matter.  
By following the neutrino-radiation hydrodynamics, 
we revealed that hyperons appear off center 
at first and prevail also at center just 
before the black hole formation occurs 
from the ever accreting proto-neutron star.  
The resulting neutrino emissions are quite similar 
to those for the purely nucleonic case 
and differs only in the earlier termination of neutrino burst.  
Once hyperons appear during the contraction of proto-neutron star 
through the mass accretion, 
they soon trigger the gravitational instability by lowering 
the critical mass and give rise to even more rapid increases of 
neutrino energies and luminosities, which are then terminated 
at the black hole formation.  

In order to claim conclusively the appearance of hyperons from observed neutrino 
signals, further systematic studies must be done.  
One must discriminate it from other possibilities such as 
the phase transitions to meson condensations 
and softer EOSs of nucleonic matter, which may produce similar short neutrino bursts.  
We remark in passing that the appearance of
quarks may not change significantly the duration of neutrino burst 
\citep{nak08} while the density profile of progenitors may affect 
the behavior of accretion luminosities \citep{sum08}.  
The prediction of event rates at neutrino detector facilities 
with the neutrino oscillations during the propagation being taken into account 
is currently under way for the models studied so far.  

The current numerical results suggest 
that the appearance of whatever new degrees of freedom 
in the black hole forming collapse will be reflected 
in the energetic neutrino signals that might be 
detected as a disappearance of massive stars in nearby galaxies 
in the planned monitoring survey \citep{koc08} 
or in large neutrino facilities \citep{and05}.  
It will be also interesting to investigate 
the hyperon emergence in the black hole formations in gamma ray bursts 
and neutron star mergers.  

\acknowledgments

The numerical simulations were performed at 
CfCA in NAOJ, JAEA,
YITP (Kyoto U.) and RCNP (Osaka U.).  
This work is partially supported by the Grants-in-Aid for the 
Scientific Research (18540291, 18540295, 19104006, 19540252) 
of the MEXT of Japan, Academic Frontier Project of the MEXT, KEK LSSP (07-05)
and the 21st-Century COE Program "Holistic Research and Education 
Center for Physics of Self-organization Systems" in Waseda University.  






\clearpage

\end{document}